\documentclass[a4paper,11pt]{article}
\usepackage{amsfonts,amsmath,amsthm,amssymb}
\usepackage{epsfig}
\begin{document}
\title{Marginal CFT perturbations at the\\ integer quantum Hall transition}
\author{Martin R. Zirnbauer\\
Institute for Theoretical Physics, University of Cologne,\\
Z\"ulpicher Stra{\ss}e 77a, 50937 K\"oln, Germany}
\date{May 31, 2021}

\maketitle

\begin{abstract}
According to recent arguments by the author, the conformal field theory (CFT) describing the scaling limit of the integer quantum Hall plateau transition is a deformed level-4 Wess-Zumino-Novikov-Witten model with Riemannian target space inside a complex Lie supergroup $\mathrm{GL}$. After a summary of that proposal and some of its predictions, the leading irrelevant and relevant perturbations of the proposed CFT are discussed. Argued to be marginal, these result in a non-standard renormalization group (RG) flow near criticality, which calls for modified finite-size scaling analysis and may explain the long-standing inability of numerical work to reach agreement on the values of critical exponents. The technique of operator product expansion is used to compute the RG-beta functions up to cubic order in the couplings. The mean value of the dissipative conductance at the RG-fixed point is calculated for a cylinder geometry with any aspect ratio.
\end{abstract}


\section{Introduction}

The integer quantum Hall (IQH) transition is the critical phenomenon that takes place when a two-dimensional electron gas in the integer quantum Hall regime is driven from one Hall plateau to the next one by varying, e.g., a magnetic field. With disorder playing a major role, the IQH transition belongs to the class of Anderson localization-delocalization transitions and as such holds the potential to serve as a paradigm for Anderson transitions \cite {EversMirlin} in low dimension and between different phases of topological matter.

The study of critical properties at the IQH transition has a long history. Based on the so-called
Pruisken-Khmelnitskii two-parameter scaling hypothesis \cite{Pruisken83,Khmelnitskii83}, many numerical studies have been undertaken (a small selection of references is {\cite{Huckestein, SlevinOhtsuki, AMSD, OGE, ZWBW, SDMG}}) in order to determine the critical exponents for the universal scaling laws governing the transition. Yet, all the numerical work failed to give a consistent and convincing picture and, in spite of considerable efforts that were expended over the years, an analytical understanding of what exactly happens at the critical point also remained elusive.

The latter situation changed recently, when the author developed \cite{CFT-IQHT} a conformal field theory description of the scaling limit of the IQH transition. In that quest, several conceptual difficulties had to be overcome, especially some serious obstructions that seemed to rule out any known scenario with a current algebra as derived from a Wess-Zumino-Novikov-Witten (WZW) model. The proposal of \cite{CFT-IQHT} shows how the postulate of holomorphic factorization of the algebra of critical currents does come true by a novel mechanism of spontaneous symmetry breaking, and it predicts a fixed-point conductivity $\sigma_{xx}^\ast = 2/\pi$ (in units of $e^2/h$) and a multifractality spectrum $\Delta_q = q(1-q)/4$ consistent with known numerical results. Moreover, the proposed CFT is a deformed WZW model with the unusual property of admitting several marginal perturbations, which might be expected to correspond to the couplings $\sigma_{xy} - \sigma_{xy}^\ast$ and $\sigma_{xx} - \sigma_{xx}^\ast$ of the two-parameter scaling picture. If so, the marginality implies that the values of the critical exponents $1/\nu$ and $y$ measured in numerical experiments are actually ill-defined (they vanish in the true scaling limit of infinite system size), and the renormalization group (RG) flow near the fixed point is logarithmically slow.

To add some perspective to our work, we remark that WZW models had already been discussed 20 years ago as possible field theories of 2D Anderson delocalized criticality. Yet, at that time the consensus was that WZW models suffer from relevant perturbations that could only be avoided by fine-tuning of the disorder, and hence their significance for real physical systems was left in doubt. More recently, however, WZW models for disordered electrons were promoted to physical relevance in the discovery \cite{SRFL-2008} of some 3D topological phases of matter, by their role as RG-stable models describing the 2D surface states of bulk topological insulators and superconductors in the symmetry classes $A{\rm I\! I\! I}$, $C{\rm I}$, and $D{\rm I\!I\!I}$ \cite{FosterEtAl2014,Foster-PRX2020}. These developments are reviewed in \cite{FosterReview}. We note that the WZW model to be discussed here is on a different footing: it is stabilized not by topological protection, but rather by spontaneous symmetry breaking and rank reduction due to strong disorder.

The plan of the paper is as follows. We begin in Section \ref{sect:scaling} with the conventional scaling hypothesis and its failure as a framework for past numerical work to produce mutually consistent results. In Section \ref{sect:review}, we summarize the conformal field theory proposed in \cite{CFT-IQHT}, review the operator product expansions that follow from its fixed-point Lagrangian, and list a few predictions of the theory. The main part of the paper is Section \ref{sect:margin}, where we investigate the possible candidates for marginal CFT perturbations and use the method of operator product expansion to compute the RG beta functions for what might be the marginally irrelevant flow into the fixed point. Section \ref{sect:av-G} calculates the disorder average of the dissipative conductance for a cylinder of length $L$ and circumference $W$, in the scaling limit $L \to \infty$ (keeping $W/L$ fixed) at criticality.

\section{Scaling hypothesis}\label{sect:scaling}

The first attempt to gain an analytical understanding of the integer quantum Hall transition was made by Pruisken and collaborators \cite{Pruisken83}. Building on the theory of weak localization for metals in a weak magnetic field, they wrote down a field-theory Lagrangian that has the dissipative conductivity $\sigma_{xx}$ and the Hall conductivity $\sigma_{xy}$ for its two coupling parameters:
\begin{equation}
    \mathcal{L} = \frac{\sigma_{xx}}{8} \, \delta^{\mu\nu} \, \mathrm{STr}\,  \partial_\mu Q\, \partial_\nu Q + \frac{\sigma_{xy}}{8}\, \varepsilon^{\mu\nu}\, \mathrm{STr}\, Q \partial_\mu Q\, \partial_\nu Q .
\end{equation}
The field $Q$ of that so-called nonlinear sigma model is a matrix which can be diagonalized with two eigenvalues, $+1$ and $-1$:
\begin{equation}
    Q = u \Sigma_3 u^{-1} , \quad \Sigma_3 = \begin{pmatrix} +\mathbf{1} &0 \cr 0 &- \mathbf{1} \end{pmatrix} .
\end{equation}
(Note that this is relevant information in view of our assertion \cite{CFT-IQHT} that the diagonalizability of $Q$ is actually lost at the RG-fixed point for the critical system.) The precise nature of the matrix $Q$ depends on whether one assumes the replica trick with bosonic or fermionic fields or employs the Wegner-Efetov supersymmetry method \cite{Efetov-Book,Wegner-Book}. Here we use the latter, where $\mathrm{STr} \equiv \mathrm{Tr}_{\rm even} - \mathrm{Tr}_{\rm odd}$ stands for the supertrace, and $u \in U \equiv \mathrm{U}(r,r|2r)$, with $r$ the number of field copies needed to express a given observable of interest.

Assuming the renormalizability of Pruisken's Lagrangian in conjunction with what should happen in limiting cases, the model's RG flow was conjectured by Khmelnitskii \cite{Khmelnitskii83}. Its central feature is an RG-fixed point at $\sigma_{xy} = 1/2$ (in units of $e^2 / h$) and an unknown value of $\sigma_{xx}\,$. Now while this so-called Pruisken-Khmelnitskii scaling picture of the plateau transition inspired a lot of activity, it has to be said that the Pruisken Lagrangian was never derived in a mathematically controlled manner. In fact, while Pruisken's derivation goes through in a weak magnetic field, it fails in the lowest Landau level, and so does the author's derivation \cite{MRZ-network} applying the color-flavor transformation to the Chalker-Coddington network model \cite{Chalker} with a single channel. Anyway, the RG-fixed point of Pruisken's model (if such a fixed point exists) lies deep in the strong-coupling regime, so it is not surprising that the model never led to any quantitative predictions about the critical behavior.

Nonetheless, Pruisken's model was construed to underpin the general belief that the renormalization group beta functions (giving the RG flow as a function of the running short-distance cutoff $a$) are of standard form near the fixed point:
\begin{equation}\label{eq:linBeta}
    \frac{d\, \sigma_{xx}}{d\ln a} = y\, (\sigma_{xx} - \sigma_{xx}^\ast) + ... \,, \quad \frac{d\, \sigma_{xy}}{d\ln a} = \nu^{-1} (\sigma_{xy} - \sigma_{xy}^\ast) + ... \,,
\end{equation}
leading to standard critical behavior (with $y < 0 < \nu^{-1}$), in particular a power-law divergence of the localization length ($\xi \sim |E - E_{\rm c}|^{-\nu}$) near criticality. Based on that belief, many authors have calculated the critical indices $\nu$ and $y$ by numerical simulation of various models --- we will not go into any details here but simply offer an incomplete list of selected references: {\cite{Huckestein, SlevinOhtsuki, AMSD, OGE, ZWBW, SDMG}}. A short synopsis is that the published results for $y$ lie in the range of $-1 < y \leq 0$ and for $\nu$ between  $2.34$ and $2.61$ (be informed, however, that a very recent study of a two-channel network model by Dresselhaus, Sbierski, and Gruzberg \cite{DSG} finds values for $\nu$ as large as $3.4$ and even $3.9$).

To explain the discrepancies between all these results, claims have been made that there exists more than one RG-fixed point and hence more than one universality class. In fact, a recent paper \cite{KNS} goes as far as suggesting that there might be a whole \emph{line} of fixed points depending on a certain parameter $p$. In the opinion of the present author, there is no justification for such claims, as they are in conflict with the Harris criterion \cite{Harris-crit}, a well-established and non-perturbative result constraining disordered criticality.

The main purpose of the present paper is to investigate another scenario that has been put forward \cite{CFT-IQHT} to explain the discord: the numerical data were analyzed with assumptions that turn out to be incorrect. Basic to our explanation is that the conformal field theory (reviewed in the next section), which emerges as the RG-fixed point, has the following property: its least irrelevant perturbation is marginal, and its most relevant perturbation is also marginal! This means that the RG flow vanishes in the linear approximation for the field-theory couplings (proportional, presumably, to $\sigma_{xx} - \sigma_{xx}^\ast$ and $\sigma_{xy} - \sigma_{xy}^\ast$); it becomes nontrivial only when nonlinear terms are taken into account. According to that scenario, the exponents $y$ and $1/\nu$ measured in numerical experiments are not well-defined (they vanish upon true extrapolation to the asymptotic scaling limit) and, therefore, numerical calculations done in the past failed to give a consistent picture.

\section{CFT of RG-fixed point}\label{sect:review}

To describe the scaling limit of the integer quantum Hall transition, this author in \cite{CFT-IQHT} proposed a conformal field theory (CFT), which is to be summarized and reviewed in the present section. It should be stressed right away that our theory is not directly related to Pruisken's model. Indeed, the latter is supposed to come from a generalized Hubbard-Stratonovich transformation and its field ($Q$) has the physical dimension of frequency, whereas the former is derived via non-Abelian bosonization \cite{Witten} and its field ($M$) carries the physical dimension of time or inverse frequency. The full derivation of our CFT requires novel techniques (transcending bosonization) and will be presented elsewhere. A feature worth highlighting is that the CFT target space resides inside a nilpotent adjoint orbit ($Q = u Q_0 u^{-1}$ and $Q_0^2 = 0$) of the global symmetry group $\mathrm{U}(r,r|2r)$, following a novel scenario of spontaneous symmetry breaking and rank reduction ($2r \to r$).

\subsection{Deformed WZW model}

As put forward in \cite{CFT-IQHT}, the conformally invariant fixed-point theory for the integer quantum Hall transition is a deformed Wess-Zumino-Novikov-Witten (WZW) model of fields $M :\; \Sigma \to X$. Its target space $X$ is of type $A|A$ \cite{suprev}, which means that the supermanifold $X$ complexifies to the complex Lie supergroup $X_\mathbb{C} = \mathrm{GL} (r|r)$ and the base is a direct product $X_0 \times X_1$ of the positive Hermitian matrices $X_0 \equiv \mathrm{Herm}^+(r)$ with the unitary matrices $X_1 \equiv \mathrm{U}(r)$, where $r$ is the number of replicas. The standard WZW action functional is
\begin{equation}\label{eq:WZW-Sn}
    S_n^{\rm WZW}[M] = \frac{\mathrm{i}n}{4\pi} \int_\Sigma \left( \mathrm{STr}\, M^{-1} \partial M \wedge M^{-1} \bar\partial M + {\textstyle{\frac{1}{3}}} d^{-1} \mathrm{STr}\, (M^{-1} dM)^{\wedge 3} \right) .
\end{equation}
Here $d = \partial + \bar\partial$, and $\partial = dz\, \partial_z\,$, $\bar\partial = d\bar{z}\, \partial_{\bar{z}}$ if $z,\bar{z}$ are complex coordinates for the Riemann surface $\Sigma$, and the notation $d^{-1} \omega$ means any two-form potential of the closed three-form $\omega$. For our case, where $\omega = \mathrm{STr}\, (M^{-1} dM)^{\wedge 3}$, such a potential exists only locally when $r \geq 2$. With the choice of normalization made, the resulting ambiguity forces the so-called level $n$ (in the literature often denoted by $k$) to be an integer.

To match the phenomenology of the integer quantum Hall transition (and, in particular, to arrange for the fundamental field $M$ to have vanishing scaling dimension), one needs to deform \cite{CFT-IQHT} the standard functional (\ref{eq:WZW-Sn}) by a truly marginal perturbation with coupling parameter $\gamma :$
\begin{equation}\label{eq:def-actn}
    S_{n,\gamma}^{\rm WZW}[M] = S_n^{\rm WZW}[M] - \frac{\mathrm{i} \gamma}{4\pi} \int_\Sigma \mathrm{STr}\, (M^{-1} \partial M) \wedge \mathrm{STr}\, (M^{-1} \bar\partial M) ,
\end{equation}
often called the Gade-Wegner term \cite{GadeWegner} in the disordered electron literature. In the present paper we will be concerned for the most part with perturbations to the fixed-point theory (\ref{eq:def-actn}). These were addressed cursorily in \cite{CFT-IQHT} but call for more attention.

\subsection{Undeformed theory: conserved currents}

We begin by reviewing some key properties of the theory (\ref{eq:WZW-Sn}) before deformation. Foremost among these is an invariance \cite{KZ} under left and right translations,
\begin{equation}\label{eq:chiral-sym}
    M(z,\bar{z}) \mapsto g_L(z) \, M(z,\bar{z})\,  g_R(\bar{z})^{-1} ,
\end{equation}
where the mapping $g_L :\; \Sigma \to \mathrm{GL}(r|r)$ is holomorphic and $g_R$ is anti-holomorphic. The left translations generate a current algebra $\widehat{\mathfrak{gl}} (r|r)_n$ with holomorphic conserved current ($\bar\partial J = 0$) defined by
\begin{equation}\label{eq:def-J}
     \frac{d}{dt} \Big\vert_{t=0} S_{n}^{\rm WZW}[\mathrm{e}^{-tY} M] = \frac{1}{2\pi \mathrm{i}} \int_\Sigma \mathrm{STr} \, J \wedge \bar\partial Y .
\end{equation}
Inserting $\mathrm{e}^{-tY} M$ into the argument of the action functional $S_{n}^{\rm WZW}$ given in Eq.\ (\ref{eq:WZW-Sn}) and linearizing in $t Y$, one finds
\begin{equation}\label{eq:holoJ}
    J = n\, \partial M \cdot M^{-1} .
\end{equation}
Similarly, the right translations $M \mapsto M g_R^{-1}$ generate another copy of the same current algebra, with anti-holomorphic conserved current ($\partial \bar{J} = 0$)
\begin{equation}\label{eq:antiholJ}
    \bar{J} = - n\, M^{-1} \bar\partial M .
\end{equation}
The latter current is defined by
\begin{equation}\label{eq:def-barJ}
     \frac{d}{dt} \Big\vert_{t=0} S_{n}^{\rm WZW}[M \mathrm{e}^{+tY}] = \frac{1}{2\pi \mathrm{i}} \int_\Sigma \mathrm{STr} \, \partial Y \wedge \bar{J} .
\end{equation}
For later use, we note that $J$ and $\bar{J}$ get exchanged by the transformation
\begin{equation}\label{eq:parity}
    \partial \leftrightarrow \bar\partial , \quad M \leftrightarrow M^{-1} ,
\end{equation}
which is referred to as the ``parity'' symmetry of the CFT (\ref{eq:WZW-Sn}).

An overarching principle of conformal field theory with continuous internal symmetries is holomorphic factorization of the algebra of conserved currents. That principle is borne out \cite{Witten-holofact} for the WZW model (\ref{eq:WZW-Sn}), thanks to the independent left and right actions of the symmetry group. In contrast, holomorphic factorization is not known to occur for Pruisken's model.

\subsection{Operator product expansions}

In WZW models, one can in principle compute everything from the field-theory Lagrangian, with a major reason being that one has a direct handle on the operator product expansions. A few details are reviewed below. We adopt the convention to write $J \equiv J(z) \, dz$, i.e., $J$ stands for the one-form of the current while $J(z)$ denotes the coefficient matrix of the current one-form in the coordinates $z, \bar{z}$.

One starts from the fact that the holomorphic current $J$, by its very definition from Eq.\ (\ref{eq:def-J}), generates left translations of the fundamental field $M$. That property fixes the operator product expansion (OPE) of $J$ with $M:$
\begin{equation}\label{eq:OPE-JM}
    J_{\;\beta}^\alpha (z) M_{\;\delta}^\gamma (w,\bar{w}) = - (-1)^{|\beta|} \delta_{\beta}^\gamma \; \frac{M_{\;\delta}^\alpha (w,\bar{w})}{z-w} + \ldots \,.
\end{equation}
Here $|\beta|$ denotes the fermion degree, i.e.\ $|\beta| = 0$ and $|\beta| = 1$ for bosonic and fermionic indices $\beta$, respectively. The ellipses in (\ref{eq:OPE-JM}) stand for terms which are finite or zero in the limit of coinciding points ($z \to w$). Now for the purpose of book keeping, let us introduce constant parameter supermatrices $A$, $B$ and set
\begin{equation}
    J^A(z) \equiv \mathrm{STr}\, \big(A J(z) \big) .
\end{equation}
Then Eq.\ (\ref{eq:OPE-JM}) can be equivalently written in the index-free form
\begin{equation}\label{eq:OPE-JMinv}
    J^A (z) \cdot \mathrm{STr}\, M(w,\bar{w}) B = - \mathrm{STr} \, A \frac{M(w,\bar{w})}{z-w} B + \ldots \,.
\end{equation}
Similarly, one has
\begin{equation}\label{eq:OPE-JbM}
    \bar{J}^A (\bar{z}) \cdot \mathrm{STr}\, B M(w,\bar{w}) = + \mathrm{STr} \, B \frac{M(w,\bar{w})}{\bar{z}-\bar{w}} A + \ldots \,.
\end{equation}

By applying the OPE (\ref{eq:OPE-JMinv}) to the expression (\ref{eq:holoJ}) for $J$ in terms of $M$, one obtains the OPE of the current with itself:
\begin{equation}\label{eq:OPE-JJ}
    J^A(z) J^B(w) = - n\, \frac{\mathrm{STr}\, AB}{(z-w)^2} + \frac{J^{[A,B]}(w)}{z-w} + \ldots ,
\end{equation}
which features a leading singular term proportional to the quantized level $n$. The very same OPE (with $J \to \bar{J}$, $z \to \bar{z}$, and $w \to \bar{w}$) holds for the anti-holomorphic current.

Next, by implementing the CFT principle that the holomorphic current $J$ be a primary field of conformal dimension $1$ for the Virasoro algebra generated by the holomorphic part, $T(z)$, of the stress-energy tensor, i.e.,
\begin{equation}
    T(z) J^A(w) = \frac{J^A(w)}{(z-w)^2} + \frac{\partial_w J^A(w)}{z-w} + \ldots ,
\end{equation}
one uses (\ref{eq:OPE-JJ}) to find an exact expression for the latter:
\begin{equation}\label{eq:suga}
    T(z) \equiv T_{\widehat{\mathfrak{gl}} (r|r)_{n,\gamma}} = \frac{(-1)^{|\alpha| +1}}{2n} \, J_{\;\beta}^\alpha (z) J_{\;\alpha}^\beta (z) + \frac{(-1)^{|\alpha| + |\beta|}}{2n^2} \, J_{\;\alpha}^\alpha (z) J_{\;\beta}^\beta (z) .
\end{equation}
Finally, by using Eq.\ (\ref{eq:suga}),  the OPE (\ref{eq:OPE-JM}), and the associativity of the operator algebra, one infers how $M$ transforms under the conformal group generated by $T(z)$:
\begin{equation}\label{eq:OPE-TM}
    T(z) M(w,\bar{w}) = h \, \frac{M(w,\bar{w})}{(z-w)^2} + \frac{\partial_w M(w,\bar{w})}{z-w} + \ldots, \quad h = \frac{1}{2n^2} \,.
\end{equation}
Here we omitted the indices of the supermatrix $M(w,\bar{w})$. Since the current algebra for $\bar{J}$ is the same, all the same formulas (with $z \to \bar{z}$, etc.) hold on the anti-holomorphic side.

To conclude this subsection, let us draw attention to a phenomenon of \emph{rank reduction}: the current algebra at work, namely $\widehat{ \mathfrak{gl}} (r|r)_{n}\,$, has a rank which is only \emph{half} of that of the global symmetry group $U = \mathrm{U}(r,r|2r)$.

\subsection{Deformed theory}

\subsubsection{Conserved currents}\label{sect:JLJR}

We now turn on the deformation in (\ref{eq:def-actn}) with parameter $\gamma$. The first question to address is what happens to the conserved currents. These are still defined by the continuous symmetry under left and right translations:
\begin{equation}
     \frac{d}{dt} \Big\vert_{t=0} S_{n,\gamma}^{\rm WZW}[\mathrm{e}^{-tX_L} M \mathrm{e}^{tX_R}] = \frac{1}{2\pi \mathrm{i}} \int_\Sigma \big( \mathrm{STr} \, J_L \wedge d X_L - \mathrm{STr}\, J_R \wedge d X_R \big) .
\end{equation}
By computing the additional terms due to the deformation (\ref{eq:def-actn}), one finds the expression
\begin{equation}\label{eq:JL}
    J_L = J - \gamma\, \mathrm{STr}\, (M^{-1} \partial M) \times \mathbf{1} + \frac{\gamma}{2}\, d \, \mathrm{STr} \ln M
\end{equation}
for the deformed current due to left translations and
\begin{equation}\label{eq:JR}
    J_R = \bar{J} + \gamma \, \mathrm{STr}\, (M^{-1} \bar\partial M) \times \mathbf{1} - \frac{\gamma}{2}\, d \, \mathrm{STr} \ln M
\end{equation}
for the deformed right current. Here one needs to appreciate the appearance of an exact term, $d\, \mathrm{STr} \ln M$. Omitted in \cite{CFT-IQHT}, that term is negligible for certain purposes: it drops out of the total current $J_L + J_R$ (from which the physical observables are computed), and the equations of motion ($d J_L = 0 = d J_R$) imply that the deformed currents with the exact term omitted are still holomorphic resp.\ anti-holomorphic, giving rise to good current algebras. Nonetheless, the exact term does affect some of the operator product expansions of the deformed CFT, and we shall therefore be careful \underline{not} to omit it here.

Let us add the remark that the deformation (\ref{eq:def-actn}) preserves the symmetry under parity, (\ref{eq:parity}). By consequence, parity still exchanges the deformed currents ($J_L \leftrightarrow J_R$).

\subsubsection{Marginality of deformation}\label{sect:gam-marg}

The main goal of the present paper is to compute the renormalization group flow of the deformed theory (\ref{eq:def-actn}) in the presence of various perturbations of physical interest. For that task, our main goal is the method of operator product expansion. Now as we have remarked, the appearance of the exact term $d\, \mathrm{STr} \ln M$ in the expressions (\ref{eq:JL}, \ref{eq:JR}) for the deformed currents has an effect on operator products. That is a nuisance when doing perturbative calculations of higher order and presents us with a dilemma. Our resolution is to do perturbation theory around the undeformed WZW action functional of Eq.\ (\ref{eq:WZW-Sn}). In other words, we shall lump all of the deformation in (\ref{eq:def-actn}) with whatever perturbation is being considered. Such an approach would be unlikely to be manageable for a general deformation. However, the deformation at hand is an Abelian current-current perturbation, which makes for substantial simplifications.

Most importantly, the deformation (\ref{eq:def-actn}) is a \emph{truly marginal} perturbation of the CFT (\ref{eq:WZW-Sn}), i.e.\ the coupling $\gamma$ does not flow under renormalization. Let us briefly discuss this fact, introducing along the way some notation that will be useful later. We abbreviate
\begin{equation}\label{eq:def-OA}
    \mathrm{STr} J(z) \, \mathrm{STr} \bar{J}(\bar{z}) \equiv O_A (z,\bar{z}) ,
\end{equation}
to cast the deformation in (\ref{eq:def-actn}) in the form
\begin{equation}\label{eq:conv-Sg}
    - \frac{\mathrm{i} \gamma}{4\pi} \int_\Sigma \mathrm{STr}\, (M^{-1} \partial M) \wedge \mathrm{STr}\, (M^{-1} \bar\partial M) = \frac{\gamma}{2\pi n^2} \int d^2z \, O_A(z,\bar{z}) \equiv S_\gamma \,,
\end{equation}
where $d^2 z \equiv |dx \wedge dy|$ stands for the area element of $\Sigma$.

It is clear that our perturbation (\ref{eq:conv-Sg}) is marginal in linear order. Indeed, being a product of holomorphic and anti-holomorphic currents, $O_A$ is a Virasoro-primary field with conformal dimension (1,1), so its integral $\int d^2 z\, O_A$ has total scaling dimension $1 + 1 - 2 = 0$. To go beyond the linear approximation, we expand:
\begin{equation}\label{eq:expand-Sg}
    \mathrm{e}^{-S_\gamma} = 1 - S_\gamma + \frac{1}{2!} S_\gamma^2
    - \frac{1}{3!} S_\gamma^3 + \ldots \,.
\end{equation}
The simplifying feature here is that, by Eq.\ (\ref{eq:OPE-JJ}) and $\mathrm{STr}\, \mathbf{1} = 0$ and $[\mathbf{1},\mathbf{1}] = 0$, the OPE
\begin{equation}
    \mathrm{STr}\, J(z) \cdot \mathrm{STr}\, J(w) = - n\, \frac{\mathrm{STr}\, \mathbf{1}^2}{(z-w)^2} + \frac{J^{[\mathbf{1},\mathbf{1}]}(w)}{z-w} + \ldots = 0 + \ldots ,
\end{equation}
contains no singular terms (the same holds, of course, on the anti-holomorphic side), and hence the limit
\begin{equation}
    \lim_{z \to w} O_A(z,\bar{z}) \, O_A(w,\bar{w}) = \; :\! O_A(w,\bar{w})^2 \!:
\end{equation}
is \emph{finite}, and so are all operator products in the expansion (\ref{eq:expand-Sg}). By standard reasoning (which will be explicated in much detail below), the true marginality of $S_\gamma$ follows.

\subsubsection{Zero scaling dimension of $M$}\label{sect:scaldimM}

The physical meaning of our WZW field $M$ is that represents the local density of states (or the absolute value squared of a critical wave function \cite{CFT-IQHT}) of the underlying disordered electron model. From that correspondence, one knows that $M$ must have vanishing scaling dimension. We are now going to demonstrate that this requirement fixes the deformation parameter to be $\gamma = 1$.

To begin, we recall from the OPE (\ref{eq:OPE-TM}) that $M$ is a Virasoro-primary field with conformal dimension $(1/2n^2,1/2n^2)$ in the undeformed theory (\ref{eq:WZW-Sn}). This means that under a change of cutoff scale $a \to a^\prime $, the local field $M(z,\bar{z})$ (indices omitted) renormalizes as
\begin{equation}
    M(z,\bar{z}) \to \left( a^\prime / a \right)^{-1/n^2} M(z,\bar{z}) .
\end{equation}
Our task is to compute the modification of this law due to the CFT deformation $S_\gamma$. In preparation for that, we apply Eqs.\ (\ref{eq:OPE-JMinv}, \ref{eq:OPE-JbM}) to compute the OPE with $O_A(z,\bar{z})$:
\begin{equation}\label{eq:OPE-easy}
    O_A(z,\bar{z}) \cdot M(0) = - \frac{1}{|z|^2} M(0) + \ldots \,.
\end{equation}
Thus, for the linear term in the Taylor series (\ref{eq:expand-Sg}) we have
\begin{equation}
    - S_\gamma \cdot M(0) = \frac{\gamma}{2\pi n^2} \int \frac{d^2z}{|z|^2}\, M(0) + \ldots \,.
\end{equation}
We then see that raising the cutoff $a \to a^\prime$ has the following effect:
\begin{equation}\label{eq:RG-M0}
    - S_\gamma \cdot M(0) \Big\vert_a + S_\gamma \cdot M(0) \Big\vert_{a^\prime}
     = \frac{\gamma}{n^2} \ln(a^\prime / a) \, M(0) + \ldots \,.
\end{equation}

Owing to the simplicity of the OPE (\ref{eq:OPE-easy}), it is straightforward to extend this calculation to all orders in $\gamma\,$; alternatively, we may take the increment in $a^\prime / a = (a + da) / a = 1 + d\ln a$ to be infinitesimal and stop at the linear order (\ref{eq:RG-M0}) to derive the differential equation that determines the renormalization of the local field $M(0)$. Either way, we find that the conformal dimension of $M$ (in the deformed CFT with parameter $\gamma$) is exactly given by
\begin{equation}
    (h,\bar{h}) = \left( \frac{1-\gamma}{2n^2} , \frac{1-\gamma}{2n^2} \right) .
\end{equation}
Since the phenomenology of IQHT calls for $h = \bar{h} = 0$, it follows that $\gamma = 1$.

\subsubsection{Some predictions at criticality}

Beyond reproducing known results, the CFT (\ref{eq:def-actn}) predicts many new ones, which can be verified by numerical simulation of critical models. Here we mention just a few.

Firstly, an observable that has received much attention {\cite{EversEtAl, ObuseEtAl, BWZ1, BWZ2}} is the spectrum of multifractal scaling exponents, $\Delta_q\,$, defined as the scaling dimensions of powers $|\psi(\bullet)|^{2q}$ of critical wave functions. These can be calculated in the present theory with a single replica ($r = 1$), in which case $M$ is simply a $2 \times 2$ supermatrix:
\begin{equation}
    M = \begin{pmatrix} M_{\;0}^0 &M_{\;1}^0 \cr M_{\;0}^1 &M_{\;1}^1 \end{pmatrix} .
\end{equation}
The wave function observable $|\psi|^{2q}$ for a microscopic model corresponds to powers $(M_{\;0}^0)^q$ of the local field $M_{\;0}^0$ in the CFT. By an easy extension of the argument given in the preceding subsection, one shows that the total scaling dimension of $(M_{\;0}^0)^q$ is
\begin{equation}
    \Delta_q = \frac{q(1-q)}{n} \quad (n \in \mathbb{N}) .
\end{equation}
By matching to the existing numerical results for $\Delta_q$ one infers that $n = 4$.

Secondly, the operator product expansion (\ref{eq:OPE-JJ}) for the current algebra $\widehat{ \mathfrak{gl}}(r|r)_{n}$ gives access to several current-current correlation functions of physical interest, at criticality. Foremost among these is the correlation function between derivatives of retarded and advanced single-electron Green's functions,
\begin{equation}\label{eq:KGref}
    \langle (\stackrel{\!\!\!\!\rightarrow}{\nabla_x} -
    \stackrel{\!\!\!\!\leftarrow}{\nabla_x}) G^{\rm ret} (\bullet,\bullet^\prime) (\stackrel{\!\!\!\!\!\rightarrow}{\nabla_{x^\prime}} -
    \stackrel{\!\!\!\!\!\leftarrow}{\nabla_{x^\prime}}) G^{\rm adv}(\bullet^\prime,\bullet)\rangle,
\end{equation}
which enters the Kubo-Greenwood linear-response formula for the conductivity $\sigma_{xx}\,$. That correlation function is represented in the CFT (still with a single replica, $r = 1$) by
\begin{equation}\label{eq:JJ-corr}
    \langle J_{\;0}^{1}(\bullet) J_{\;1}^{0}(\bullet^\prime) \rangle +
    \langle \bar{J}_{\ 0}^{\,1}(\bullet) \bar{J}_{\ 1}^{\,0}(\bullet^\prime) \rangle ,
\end{equation}
which has a leading short-distance singularity proportional to $n$ from (\ref{eq:OPE-JJ}). By computing the correlation function (\ref{eq:JJ-corr}) in the perturbative short-distance regime of a finite system with absorbing boundaries, one infers the value of the fixed-point dissipative conductivity as
\begin{equation}
    \sigma_{xx}^\ast = \frac{n}{2\pi} \approx 0.6367 \quad (n = 4),
\end{equation}
see Section \ref{sect:G-pert}. (See also Section \ref{sect:G-exact} for the exact calculation.)

Thirdly, it is possible to compute the mean dissipative conductance in the scaling limit of a cylinder geometry with any aspect ratio; see Section \ref{sect:av-G}.

\section{Marginal perturbations}\label{sect:margin}

In the traditional framework of two-parameter scaling theory based on Pruisken's nonlinear sigma model, one envisages two leading perturbations to the renormalization-group (RG) fixed point: a relevant one given by $\sigma_{xy} - \sigma_{xy}^\ast \,$, and an irrelevant one, by $\sigma_{xx} - \sigma_{xx}^\ast \,$. A conservative approach ought to try and rescue as much of that framework as possible. In that vein, we now look for two perturbations of the above kind for the CFT (\ref{eq:def-actn}).

\subsection{Ruling out a natural candidate}\label{sect:ruleout}

We start our investigation of perturbations of the fixed-point theory (\ref{eq:def-actn}) by considering
\begin{equation}\label{eq:1st-pert}
    S_{\rm pert} = \frac{\mathrm{i}\lambda}{2} \int \mathrm{STr} \, J \wedge M \bar{J} M^{-1} ,
\end{equation}
which would seem to be a natural candidate for our purposes. It turns out, however, that the perturbation (\ref{eq:1st-pert}) is ruled out by the phenomenology to be matched. Truth be told: that no-go scenario came as a surprise to this author; we shall now present the details.

To linear order in the coupling, the RG flow of the perturbation is controlled by the operator product expansion (OPE) with the stress-energy tensor $T(z)$. As outlined in the opening paragraph of Sect.\ \ref{sect:gam-marg}, we organize our perturbation expansion around the CFT (\ref{eq:WZW-Sn}); i.e., we take the OPE with the stress-energy tensor (\ref{eq:suga}) of the undeformed theory and account for the deformation $S_\gamma$ by the perturbative scheme of Sect.\ \ref{sect:scaldimM}.

To get started, we compute the OPE of $T(z)$ as given in Eq.\ (\ref{eq:suga}) with the integrand of $S_{\rm pert}\,$, abbreviated as
\begin{equation}\label{eq:O-M}
    O(w,\bar{w}) = \mathrm{STr}\, J(w) M(w,\bar{w}) \bar{J}(\bar{w}) M^{-1}(\bar{w},w) .
\end{equation}
Because operator product expansion is an associative process and $T(z)$ is quadratic in currents, the computation proceeds via the OPE of a current with $O(w,\bar{w})$. Thus our first step is to expand the operator product $J(z) \cdot O(w,\bar{w})$. Now since $O(w,\bar{w})$ does not separate into holomorphic and anti-holomorphic factors, that OPE is not immediately available from the results we have accumulated so far. So, in order to perform the required expansion, which is fundamental to everything we do here and in the sequel, let us take a time-out and review the basic trick \cite{KZ} behind it.

\subsubsection{Tutorial}

Consider a left translation $M \mapsto \mathrm{e}^{-tY} M$ generated by a differentiable field configuration $Y$. Use the fact that, by definition, the functional integral measure $\mathcal{D} M$ is invariant under such translations. Let $\delta_Y$ denote the differential operator defined by linearization at $t = 0$, and observe that the derivation $\delta_Y$ obeys the Leibniz product rule. Take $Y(z,\bar{z})$ to be holomorphic in the disk $D:\; |z-w| < \epsilon_1\,$, smooth in the annulus $A:\; \epsilon_1 < |z-w| < \epsilon_2\,$, and identically zero in the exterior, i.e.\ for $|z-w| > \epsilon_2\,$. Then, assuming that there are no extraneous operator insertions in the support of $Y$, which is the disk $|z-w| \leq \epsilon_2\,$, derive the validity of
\begin{equation}
    (\delta_Y O)(w,\bar{w}) = O(w,\bar{w}) \, \delta_Y S_{n}^{\rm WZW}
\end{equation}
by partial integration under the functional integral sign. Now recall the definition (\ref{eq:def-J}) of the current $J$ to verify that
\begin{equation}
    2\pi \mathrm{i} \, \delta_Y S_{n}^{\rm WZW} = \int_A \mathrm{STr}\, J \wedge \bar\partial Y = \oint_{\partial D} \mathrm{STr}\, Y J(z)\, dz ,
\end{equation}
where the area integral was converted into a line integral around the boundary $\partial D$ of the inner disk by the equation of motion $\bar\partial J = 0$ and Stokes' formula. It follows that
\begin{equation}\label{eq:OPE-formula}
    (\delta_Y O)(w,\bar{w}) = \frac{1}{2\pi\mathrm{i}} \oint_{\partial D} \mathrm{STr}\, Y J(z)\, dz \cdot O(w,\bar{w})
\end{equation}
holds under the functional integral sign. Finally, expand the holomorphic function $Y$ in the disk $D$ as
\begin{equation}
    Y(z) = Y(w) + (z-w) Y^\prime(w) + \ldots \,,
\end{equation}
insert this expansion into Eq.\ (\ref{eq:OPE-formula}), and compare the coefficients of $Y(w)$, $Y^\prime(w)$, etc., to infer the terms singular as $(z-w)^{-k}$ ($k = 1, 2, ...$) in the operator product expansion of $J(z)$ multiplying $O(w,\bar{w})$. This ends our tutorial review of the basic trick of the trade. Armed with it, we are in a position to tackle the computation of the OPE at hand.

\subsubsection{Renormalization of $O(w,\bar{w})$}

So far, we have not used any property of $O(w,\bar{w})$ other than locality. Now, we recall the specific form (\ref{eq:O-M}) of $O(0)$ (using invariance under spatial translations to set $w = \bar{w} = 0$ for notational simplicity) and compute the left-hand side of (\ref{eq:OPE-formula}) from the expression (\ref{eq:holoJ}) for the current $J$:
\begin{equation}\label{eq:varO2}
    (\delta_Y O)(0) = - n \, \mathrm{STr} \, Y^\prime(0) \big( M \bar{J} M^{-1} \big)(0) .
\end{equation}
Here no term $Y(0)$ (without prime) appears because $O$ is invariant under left translations generated by a constant $Y$. Thus we infer that the desired OPE is
\begin{equation}\label{eq:int-res}
    J_{\;\alpha}^\beta(z) \, O(0) = \frac{-n}{z^2} \, \big( M \bar{J} M^{-1} \big)_{\;\alpha}^\beta (0) + \ldots ,
\end{equation}
where a simple pole $z^{-1}$ is absent due to the absence of $Y(0)$ on the r.h.s.\ of (\ref{eq:varO2}).

Next we compute the OPE of $(-1)^{|\alpha|} J_{\;\beta}^\alpha(z)$ with the intermediate result above. The outcome can be presented as
\begin{equation}
    \mathrm{STr}\, J(z) J(z) \cdot O(0) = \frac{-2n}{z^2}\, O(0) + \ldots \,.
\end{equation}
Here, the factor of 2 comes from taking into account the finite term $:\! J_{\;\alpha}^\beta (0) \, O(0) \!:$ not displayed in (\ref{eq:int-res}). (We follow the standard convention of using colons to denote the finite part of an operator product.) Note also that there is no need for point splitting of the two currents in $T(z)$ since $\mathrm{STr} :\! J(z) J(z) \!: \; = \mathrm{STr} \, J(z) J(z)$ by Eq.\ (\ref{eq:OPE-JJ}) and $\mathrm{STr}\, \mathbf{1} = 0$.

What we have computed so far is the OPE with the first summand in the expression (\ref{eq:suga}) for $T(z)$. To complete that computation, we need to include the contribution from the second summand. Simply repeating the steps of before, we obtain
\begin{equation}\label{eq:TzO0}
    T(z) \cdot O(0) = \frac{O(0)}{z^2} - \frac{1}{n} \, \frac{O_A(0)}{z^2} + \ldots ,
\end{equation}
where $O_A$ was introduced in Eq.\ (\ref{eq:def-OA}).

It remains to account for the presence of the deforming factor $\mathrm{e}^{-S_\gamma}$. Its effect on the renormalization of $O(w,\bar{w})$ is seen to be nil, as follows. We compute the OPE with $O_A :$
\begin{eqnarray}
    &&O_A(z,\bar{z})\, O(0) = - \frac{n}{z^2} \, \big( \mathrm{STr} \bar{J}(0) \big)^2 - \frac{n}{\bar{z}^2} \, \big( \mathrm{STr} J(0) \big)^2 + :\! O_A(0) O(0) \!: + \ldots \,.
\end{eqnarray}
There is no qualitative change when we go to higher powers of $O_A$ --- all terms in the OPE are either finite or singular as $z^{-2}$ or $\bar{z}^{-2}$. Now those singular terms average to zero by angular dependence when we take the integral $\int d^2z \, O_A(z,\bar{z})$. Thus we learn that $\mathrm{e}^{-S_\gamma} \cdot O(w,\bar{w})$ is totally finite. Therefore, the $\gamma$-deformation cannot affect (in first order) the renormalization of $S_{\rm pert} = \lambda \int d^2w \, O(w,\bar{w})$.

\subsubsection{Discussion}

What do we learn from the result above? The first term on the right-hand side of Eq.\ (\ref{eq:TzO0}) is what characterizes a marginal perturbation $\int d^2w\, O(w,\bar{w})$, with total scaling dimension $1+1-2 = 0$. Yet, there is also the second term. Its presence implies that any small but nonzero coupling $\lambda$ inevitably drives RG flow of the parameter $\gamma$ of the fixed-point action (\ref{eq:def-actn}). Normalization taken into account, we find that the RG flow equation for $\gamma$ is
\begin{equation}
    \frac{d \gamma}{d\ln a} = 4\pi n \, \lambda + ... \,,
\end{equation}
to linear order in the coupling $\lambda$. We observe that the right-hand side of this equation is independent of $\gamma$ and, in particular, $\gamma = 1$ is not a fixed point of the flow.

On the other hand, the fixed-point value $\gamma = 1$ is not disposable, as it sets the scaling dimension of $M$ to zero, cf.\ Eq.\ (\ref{eq:OPE-TM}), thereby matching a hallmark of the IQH transition. The conclusion then is that $S_{\rm pert}$ in Eq.\ (\ref{eq:1st-pert}) is \underline{not} a viable perturbation for us to add to the fixed-point action (\ref{eq:def-actn}). We remark in passing that the same result had been reached in {\cite{Guruswamy, FosterEtAl2014, XCF-2015}} from a one-loop calculation of the RG beta function.

This state of affairs begs the question: how can one explain on microscopic grounds that the dangerous coupling $\lambda$ vanishes in the perturbed CFT? Most likely, the answer is that our CFT emerges from a process of non-Abelian bosonization (where we anticipate that the theory is derived by such a step starting from a precursor theory of Dirac-type fields), and that step results in the action functional (\ref{eq:def-actn}) right at the fixed-point value for the kinetic coupling, i.e.\ with perturbation parameter $\lambda = 0$.

\subsection{Marginal current-current perturbation}\label{sect:JJ-pert}

Having learned that the perturbation (\ref{eq:1st-pert}) is ruled out by what we know about the IQH transition, we move on to investigate another possibility in the present subsection. We shall consider a perturbation
\begin{equation}\label{eq:2nd-pert}
    S_{\rm pert} = \frac{1}{2\pi n^2} \int d^2z\, ( \gamma \cdot O_A + \delta \cdot O_I )
\end{equation}
by two current-current interactions:
\begin{equation}\label{eq:OAOI}
    O_A(z,\bar{z}) = \mathrm{STr}\, J(z) \, \mathrm{STr}\, \bar{J}(\bar{z}) ,
    \quad O_I(z,\bar{z}) = \mathrm{STr}\, J(z) I \bar{J}(\bar{z}) I^{-1} .
\end{equation}
We notice that $O_A$ is still the two-trace Gade-Wegner term, which already appeared above, while the new term $O_I$ is a single-trace non-Abelian current-current perturbation that involves some isomorphism $I$ between the spaces of right movers and left movers:
\begin{equation}
    I : \; \mathbb{C}_R^{r|r} \to \mathbb{C}_L^{r|r} . 
\end{equation}
(There exists no canonical isomorphism of this kind; its appearance needs to be motivated by microscopic reasoning and our scenario \cite{CFT-IQHT} of spontaneous symmetry breaking and rank reduction.) Both couplings, $\gamma$ and $\delta$, are marginal -- the explanation for the case of $\gamma$ has already been given in Section \ref{sect:gam-marg}; the argument for $\delta$ is no different.

The goal now is to compute the RG beta functions to cubic order in the couplings $\delta$ and $\gamma$. In view of that technical challenge, we seek a computational scheme that is maximally efficient. As we already know, a natural idea is to carry out the RG procedure of ``coarse graining'' via operator product expansion. More precisely, the scheme goes as follows. (We have already been using that scheme implicitly; but now, in view of the serious work ahead, we make it more explicit.) We expand the statistical weight function $\mathrm{e}^{- S_\ast - S_{\rm pert}}$ in powers of the perturbation $S_{\rm pert}$ of the fixed-point action functional $S_\ast\,$, then we perform OPE to extract the UV-singular local terms contained in those powers, then we carry out the cutoff-raising scale transformation by using the known OPEs with the stress-energy tensor of the CFT, and finally we re-exponentiate to infer the RG flow of the couplings. That scheme is especially smooth in the present case of marginal perturbations, where the cutoff-raising scale transformation has a trivial effect.

Before we start our investigation of the perturbation (\ref{eq:2nd-pert}), let us revisit the question of how to organize the perturbation expansion to accommodate the deformation term in Eq.\ (\ref{eq:def-actn}). As we saw in Section \ref{sect:JLJR}, the deformation corrupts the holomorphic and anti-holomorphic conserved currents by the addition of an exact term, which potentially modifies the outcome of operator product expansions involving many factors. For that reason, we chose to build the perturbation theory around the undeformed fixed-point action functional of Eq.\ (\ref{eq:WZW-Sn}):
\begin{equation}\label{eq:Sast}
    S_\ast \equiv S_{n,\gamma = 0}^{\rm WZW} \,.
\end{equation}
Another aspect worth mentioning is this: in Ref.\ \cite{CFT-IQHT} the deformation term was shown to arise as a result of integrating out the Goldstone modes (with diverging stiffness) that restore a spontaneously broken symmetry. By that token, the deformation is to be regarded as a one-loop quantum correction (it is, in fact, of order $1/n$ relative to the main term). In such a situation it is not clear, at least not immediately, how to sort the terms of a loop (or $1/n$) expansion in a consistent manner. This provides us with a second motivation for our approach based on (\ref{eq:Sast}). As a final remark let us emphasize, to make sure, that the coupling $\gamma$ in Eq.\ (\ref{eq:2nd-pert}) coincides with the coupling $\gamma$ in Eq.\ (\ref{eq:def-actn}).

\subsubsection{$O_I \times O_I$}

Here we expand the operator product $O_I \times O_I$ for the field $O_I$ of the marginal perturbation $\int d^2z \, O_I$. The main tools will be the OPE (\ref{eq:OPE-JJ}) for the holomorphic current $J$, together with the analogous OPE for $\bar{J}$, which is of identical form. As before, we use a constant parameter matrix $B$ to do the book keeping of matrix indices. Then
\begin{eqnarray}
    &&\mathrm{STr}\, B J(z) \cdot O_I(0) = - \frac{n}{z^2}\,
    \mathrm{STr}\, B I \bar{J}(0) I^{-1} \cr &&+ \frac{1}{z}\, \mathrm{STr}\, [ B , I \bar{J}(0) I^{-1} ] J(0)\, + :  O_I(0)\, \mathrm{STr} \, B J(0) : + \ldots \,.
\end{eqnarray}
Now, substituting $B \to I \bar{J}(\bar{z}) I^{-1}$ we get
\begin{eqnarray}\label{eq:midterm}
    &&O_I(z,\bar{z}) \cdot O_I(0) = - \frac{n}{z^2}\, \mathrm{STr}\, \bar{J}(\bar{z}) \bar{J}(0) \cr &&+ \frac{1}{z}\, \mathrm{STr}\, [ \bar{J}(\bar{z}) , \bar{J}(0) ] I^{-1} J(0) I \, +  O_I(0)\, \mathrm{STr} \, I \bar{J}(\bar{z}) I^{-1} J(0) + \ldots \, .
\end{eqnarray}
Next we carry out the OPE for $\bar{J}(\bar{z}) \cdot \bar{J}(0)$. Then, upon taking the integral $\int d^2z \, O_I(z,\bar{z})$, the first term on the r.h.s.\ will integrate to zero. Also, the leading contribution from the third term is disconnected, $:\! O_{\rm I}(0) O_{\rm I}(0)\! : \,$, and hence will cancel (by the linked-cluster principle) when we re-exponentiate. Thus the only term of interest here is the middle one.

Now from Eq.\ (\ref{eq:OPE-JJ}) (or, rather, the exact analog thereof for $\bar{J}$) we derive the formula
\begin{eqnarray}
    &&\mathrm{STr}\, A \bar{J}(\bar{z}) B \bar{J}(0) = - \frac{n}{\bar{z}^2} \, \mathrm{STr} A \cdot \mathrm{STr} B \cr &&+ \frac{1}{\bar z}\, \big( - \mathrm{STr} A \bar{J}(0) \cdot \mathrm{STr} B + \mathrm{STr} B \bar{J}(0) \cdot \mathrm{STr} A \big) + \ldots \, .
\end{eqnarray}
Applying this expansion (with suitable identifications for $A$ and $B$) to the middle term on the right-hand side of (\ref{eq:midterm}), we arrive at
\begin{equation}
    \frac{1}{z}\, \mathrm{STr}\, [ \bar{J}(\bar{z}) , \bar{J}(0) ] I^{-1} J(0) I = \frac{2}{|z|^2}\, \mathrm{STr} J(0)\, \mathrm{STr} \bar{J}(0) + \ldots \,.
\end{equation}
For future use, we here record the following intermediate result:
\begin{eqnarray}\label{eq:intermed}
    O_I(z,\bar{z}) \cdot O_I(0) &=& - \frac{n}{z^2}\, \mathrm{STr} :\! \bar{J}(0) \bar{J}(0)\!: - \frac{n}{\bar{z}^2}\, \mathrm{STr} :\! J(0) J(0)\!: \cr &+& :\! O_I(0) O_I(0)\!: + \frac{2}{|z|^2}\, O_A(0) + \ldots \, .
\end{eqnarray}
{}From it, we read off the UV-singular contribution:
\begin{equation}
    \int d^2z \, O_I(z,\bar{z}) \cdot O_I(0) = 2 \int \frac{d^2z}{|z|^2} \, O_A(0) + \ldots \, .
\end{equation}
Now, before we can write down the ensuing RG flow equation for the coupling $\gamma$, we need to compute another OPE giving a contribution of the same order.

\subsubsection{$O_I \times O_I \times O_A$}

In this subsection we compute the OPE for a triple product
\begin{equation}
    \int d^2u \, O_I(u,\bar{u}) \cdot \int d^2v \, O_I(v,\bar{v}) \cdot O_A(0) = \; ?
\end{equation}
Using associativity, we first expand $O_I(v,\bar{v}) \cdot O_A(0)$; afterwards, we expand the product with $O_I(u,\bar{u})$. In so doing, we need to assume that $|u| > |v|$, so as to work inside the radius of convergence of the OPE; that can be done without the complication of distinguishing cases, as $u$ and $v$ are on the same footing (both appear as arguments of the field $O_I$) and we simply exchange $u \leftrightarrow v$ when the opposite inequality holds.

Having explained in some detail the technology in use, we now switch to a more concise style of presentation. By using nothing but the various identities given above, we find
\begin{equation}
    O_I(v,\bar{v}) \cdot O_A(0) = - \frac{n}{v^2} \big( \mathrm{STr} \bar{J}(0) \big)^2 - \frac{n}{\bar{v}^2} \big( \mathrm{STr} J(0) \big)^2 +
    :\! O_I(0) O_A(0)\! : + \ldots \,.
\end{equation}
The subsequent OPE of $O_I(u,\bar{u})$ with the first two terms on the right-hand side yields coefficient functions $(\bar{u} v)^{-2}$ and $(\bar{v} u)^{-2}$, which average to zero when carrying out the angular integrations in $\int d^2u \int d^2v$. We therefore concentrate on the OPE of $O_I(u,\bar{u})$ with the third term, $:\! O_I(0) O_A(0)\! :\,$. There are two contractions here that give a nonzero final result. In the first one, the $J(u)$ in $O_I(u,\bar{u})$ contracts with the $J(0)$ in $O_A(0)$ while the $\bar{J}(\bar{u})$ in $O_I(u,\bar{u})$ contracts with the $\bar{J}(0)$ in $O_I(0)$. The other contraction pattern is obtained by exchanging the roles of $J(u)$ and $\bar{J} (\bar{u})$. Both contraction pattern make for the same contribution, which explains the combinatorial factor of 2 in the following result:
\begin{equation}
    O_I(u,\bar{u}) :\! O_I(0) O_A(0)\! : \; = 2 \left\vert \frac{-n}{u^2} \right\vert^2 O_A(0) + \ldots \,.
\end{equation}
Hence,
\begin{equation}
    \frac{1}{2!} \left( \int d^2z \, O_I(z,\bar{z}) \right)^2 O_A(0) = 2n^2 \int\!\!\!\!\!\!\!\! \int\limits_{|u| > |v|} \frac{d^2u \, d^2v}{|u|^4} \, O_A(0) + \ldots \,.
\end{equation}
By performing (for fixed $|v|$) the integral over $u$ with domain $|u| > |v|$, we arrive at
\begin{equation}
    \frac{1}{2!} \left( \int d^2z \, O_I(z,\bar{z}) \right)^2 O_A(0) = 2 \pi n^2 \int \frac{d^2v}{|v|^2} \, O_A(0) + \ldots \,.
\end{equation}
In summary, both operator products $O_I \times O_I$ and $O_I \times O_I \times O_A$ expand to produce the term $O_A$ with coupling $\gamma$. Note also that taking further operator products with $O_A$ leads to nothing singular, as the OPE of the Abelian perturbation $O_A$ with itself is non-singular. Now, by executing the scheme laid out at the beginning of the section, we can infer the leading-order RG flow equation for the coupling $\gamma$, as follows.

With our calculations so far, we have shown that the Taylor series of $\mathrm{e}^{-S_{\rm pert}}$ contains the following UV-singular terms:
\begin{eqnarray}
    \left( \mathrm{e}^{-S_{\rm pert}} \right)_{\rm sing} &=& \frac{1}{2!} \left( \frac{- \delta}{2\pi n^2} \int d^2z\, O_I \right)^2 \left( 1 - \frac{\gamma}{2\pi n^2} \int d^2z \, O_A \right)_{\rm sing} + \ldots \cr &=& \frac{\delta^2}{(2\pi n^2)^2} (1-\gamma) \int \frac{d^2v}{|v|^2} \, \int d^2z\, O_A + \ldots \,,
\end{eqnarray}
to leading order in the couplings $\delta$ and $\gamma$. We now raise the UV-cutoff of the field theory from $a$ to $a+da$. In that process, $S_\ast$ remains fixed, and so do the marginal perturbations, especially $\int d^2z \, O_A\,$. By re-exponentiating and reading off the renormalized action, we conclude that
\begin{equation}
    \frac{\gamma(a+da) - \gamma(a)}{2\pi n^2} = \frac{\delta^2}{(2\pi n^2)^2} \, (1-\gamma) \bigg( \int_{|v| > a+da} - \int_{|v| > a} \bigg) \frac{d^2v}{|v|^2} \,,
\end{equation}
which yields the RG flow equation
\begin{equation}\label{eq:flow-g}
    \frac{d\gamma}{d \ln a} = - \frac{\delta^2}{n^2} \, (1-\gamma) + \mathcal{O}(\delta^3) .
\end{equation}
We note that $\gamma = 1$ is a fixed point of the flow.


\subsubsection{$O_I \times O_I \times O_I$}

As we have seen, the coupling $\delta$ of the marginal perturbation $\int d^2z\, O_I$ remains marginal (i.e., does not flow) up to quadratic order of the expansion in $\delta$. That situation changes when we go to cubic order in $\delta$. Following the same scheme as before, we compute the RG flow of $\delta$ by expanding the operator product
\begin{equation}\label{eq:triple-OI}
    \int d^2u \, O_I(u,\bar{u}) \int d^2v \, O_I(v,\bar{v}) \cdot O_I(0) = \; ?
\end{equation}
For that, our starting point is the OPE of $O_I(v,\bar{v})$ with $O_I(0)$ as given by Eq.\ (\ref{eq:intermed}).

Without loss, we assume that $|u| > |v|$ as before. Owing to the angular integrations in $\int d^2u \int d^2v$ only the term $:\! O_I(0) O_I(0)\!:$ on the right-hand side of Eq.\ (\ref{eq:intermed}) will give a nonzero contribution to the OPE (\ref{eq:triple-OI}). Hence we compute $O_I(u,\bar{u}) :\! O_I(0) O_I(0)\!:\,$. To obtain a connected term (not canceled by re-exponentiation and the linked-cluster principle) the $J(u)$ and $\bar{J}(\bar{u})$ in $O_I(u,\bar{u})$ must contract with \emph{different} factors of the normal-ordered product $:\! O_I(0) O_I(0)\!: \,$. By exchange symmetry, there are two such contraction patterns, giving a combinatorial factor of $2$. Each comes with a factor of $|-n/u^2|^2$, so we obtain
\begin{equation}
    O_I(u,\bar{u}) :\! O_I(0) O_I(0)\!: \; = \frac{2n^2}{|u|^4}\, O_I(0) + \ldots \,.
\end{equation}
Again, integrating over $u$ we have $\int_{|u| > |v|} d^2u \, |u|^{-4} = \pi\, |v|^{-2}$. It then follows that
\begin{eqnarray*}
    &&\frac{1}{3!} \left( \frac{-\delta}{2\pi n^2} \int d^2z\, O_I(z,\bar{z}) \right)^3 \cr
    &&= \frac{-\delta^3}{3! (2\pi n^2)^3} \int d^2w\, \int\limits_{|u-w|>|v-w|} 2\, d^2v\, d^2u \, O_I(u,\bar{u})  :\! O_I(w,\bar{w})^2 \!: + \ldots \cr &&= \frac{-\delta^3}{3! (2\pi n^2)^2} \int d^2w\, \frac{O_I(w,\bar{w})}{2\pi n^2} \int d^2v\, \frac{4 \pi n^2}{|v-w|^2} + \ldots \,.
\end{eqnarray*}
By the standard reasoning explained above, we deduce the result
\begin{equation}\label{eq:flow-d}
    \frac{d\, \delta}{d \ln a} = \frac{\delta^3}{3 n^2} \,.
\end{equation}
One may ask whether this result could be modified by a contribution of order $\delta^3 \gamma$ (where $\gamma = 1$) due to the quadruple product $O_I \times O_I \times O_I \times O_A$. The answer is no, at least not in the present order $1/n^2$ of a $1/n$ expansion. Indeed, one easily verifies by power counting and inspection that one cannot manufacture a (connected) single-trace operator $O_I$ (with coefficient of order $1/n^2$) by the OPE of $O_I \times O_I \times O_I$ with the two-trace operator $O_A\,$.

\subsubsection{Discussion}

Do the RG flow equations we have computed, namely Eq.\ (\ref{eq:flow-d}) for $\delta$ and Eq.\ (\ref{eq:flow-g}) for $\gamma$, offer a plausible scenario for stable flow into the CFT (\ref{eq:def-actn})? While our expectation was to obtain a stable flow (in the absence of parity-breaking perturbations corresponding to $\sigma_{xy} - \sigma_{xy}^\ast$), the flow along the real $\delta$-axis is seen to be \emph{unstable}! Turning it around, our CFT (\ref{eq:def-actn}) with $\delta = 0$ and $\gamma = 1$ would be a stable fixed point of Eqs.\ (\ref{eq:flow-d}, \ref{eq:flow-g}) \emph{if the coupling $\delta$ were imaginary}. What should we think of that? Given that the starting point of the Wegner-Efetov formalism is a totally oscillatory field integral, imaginary couplings in the effective action are not entirely unreasonable. However, if $\delta$ is imaginary, then it cannot be linear in the real quantity $\sigma_{xx} - \sigma_{xx}^\ast\,$. Therefore, if our stable flow scenario (with imaginary coupling $\delta$) turns out to be correct, the quantitative details of how the conductivity $\sigma_{xx}$ flows to its fixed-point value $\sigma_{xx}^\ast = n/2\pi$, still remain an open question.

Are there any alternative scenarios? One possibility might be that the stable RG flow can only be understood in an extended field-theory framework, for the following reason. As of this writing, our heuristic picture of how the CFT (\ref{eq:def-actn}) arises, starts from four Dirac species (see Section 2.2 of Ref.\ \cite{CFT-IQHT}), each of which bosonizes to a level-1 WZW theory. Residual interactions then lock the four independent WZW fields into a single level $n=4$ theory, Eq.\  (\ref{eq:def-actn}), for the collective field $M$. It is conceivable that the stable RG flow we are seeking is interwoven with that locking process. Another, more exotic possibility might be that the irrelevant perturbation is non-local.

\subsection{Relevant perturbation}\label{sect:rel-pert}

The CFT perturbations we have considered up to now preserve the symmetry under ``parity'', (\ref{eq:parity}), which is a distinctive property of the system at criticality. (Appealing to the  analogy with Pruisken's nonlinear sigma model, it is the analog of inverting the topological $\theta$-angle of that model). Notwithstanding the open questions left by Sections \ref{sect:ruleout} and \ref{sect:JJ-pert} above, all such perturbations are expected to be irrelevant. To make a relevant perturbation, one should break the parity symmetry so as to favor, say, clockwise (over counterclockwise) current circulation. We now identify such a perturbation of (\ref{eq:def-actn}).

For that purpose, we work in the Euclidean plane, $\Sigma = \mathbb{R}^2$, or some domain $\Sigma \subset \mathbb{R}^2$, with Cartesian coordinates $x, y$ and complex coordinate function $z = x + \mathrm{i}y$. To motivate the proper idea, we start from the effective action (\ref{eq:GFF}) derived in \cite{CFT-IQHT} and reviewed below, and we recall that our continuum field theory arises by taking the continuum limit of a network (or other) model with \emph{short-distance cutoff}. The existence of such a cutoff allows for the possible existence of short-distance singularities in the continuum fields $\varphi$ and $\theta$ figuring in Eq.\ (\ref{eq:GFF}) of Section \ref{sect:G-exact}. For the compact boson field $\theta :\; \Sigma \to \mathbb{R}/ 2\pi \mathbb{Z}$ these are of vortex type. For example, the $\theta$-field for a single vortex centered at the point $p$ with complex coordinate $z_p$ would be
\begin{equation}\label{eq:vort1}
    \theta = \mathrm{arg}(z-z_p) = \frac{1}{2\mathrm{i}} \ln\frac{z-z_p}{\bar{z} - \bar{z}_p}\,.
\end{equation}
By the common origin of $\varphi$ and $\theta$ from the supermatrix field $M$ --- see Eqs.\ (\ref{eq:gauss}, \ref{eq:log-det}) and note the expression $\mathrm{SDet}\, M = \mathrm{e}^{\varphi - \mathrm{i}\theta}$ --- we expect a vortex at position $p$ in the $\theta$-field to be accompanied by a Hodge-dual vortex ($\star d\theta = d\varphi$) at the same position $p$ in the $\varphi$-field:
\begin{equation}\label{eq:vort0}
    \varphi = - \ln |z-z_p| \,.
\end{equation}
What might be the physical object corresponding to such a singularity in the underlying microscopic physics? The expression for a SUSY-correlated vortex (\ref{eq:vort1}) and (\ref{eq:vort0}),
\begin{equation}\label{eq:SUSY-vort}
    \mathrm{e}^{\varphi - \mathrm{i}\theta} = |z-z_p|^{-1} \, \mathrm{e}^{-\mathrm{i}\, \mathrm{arg}(z-z_p)} = (z-z_p)^{-1} ,
\end{equation}
suggests that we see an indication of the onset of pre-localized states away from criticality.
Notice, however, that there exist three more singularity types of a similar kind:
\begin{equation}\label{eq:more-vort}
    (\bar{z} - \bar{z}_p)^{-1} , \quad z-z_p \,, \quad \bar{z} - \bar{z}_p \,.
\end{equation}

The following argument is familiar from the Kosterlitz-Thouless treatment \cite{KT1973,JMK1974} of the two-dimensional $xy$ model. To estimate the statistical significance of such singularities as (\ref{eq:SUSY-vort}, \ref{eq:more-vort}), one compares the cost in energy with the gain in entropy. For the statistical weight function given by the effective action (\ref{eq:GFF}), the energy cost of an isolated vortex (in a system with short-distance cutoff $a_{\rm UV}$ and size $a_{\rm IR}$) is
\begin{equation*}
    \triangle E = \frac{n}{8\pi} (2\pi + 2\pi) \ln(a_{\rm IR} / a_{\rm UV}) ,
\end{equation*}
with the singularities in $\theta$ and $\varphi$ contributing equally. Since the vortex can be located anywhere in the system with area $(a_{\rm IR}/a_{\rm UV})^2$, the free energy for an isolated vortex is
\begin{equation}
    f = \frac{n-4}{2}\, \ln(a_{\rm IR} / a_{\rm UV}) ,
\end{equation}
which vanishes in our case of $n = 4$.

Thus our $\theta$-$\varphi$ system (\ref{eq:GFF}) sits right at the critical point of a phase transition of Kosterlitz-Thouless type! (In the case of the $xy$-model, this would be the end point of a line of fixed points.) Then, what kind of perturbation drives the system from criticality into the localized regime? One might expect the answer to be a form of chemical potential that enhances the vortex fugacity. In the $xy$-model, the enhancement could be brought about by a sine-Gordon type of interaction. However, in our case such terms as $\mathrm{STr}\, (M + M^{-1})$ or similar are forbidden \cite{CFT-IQHT} as additions to the Lagrangian (what's available are but the currents). For another requisite, the perturbation must break parity (\ref{eq:parity}), i.e.\ the invariance under $\partial \leftrightarrow \bar\partial$ and $M \leftrightarrow M^{-1}$, which is the symmetry that distinguishes the critical manifold ($\sigma_{xy} = \sigma_{xy}^\ast$). Guided by these reasons, we propose the wanted term to be
\begin{equation}\label{eq:S-rel}
    S_{\rm pert}^+ = \frac{\delta_+}{\mathrm{i} \pi} \int_\Sigma \partial\, \mathrm{STr}\, M^{-1} \bar\partial M .
\end{equation}
To verify that this perturbation does do the job of breaking parity, note that while the two-form integrand
\begin{equation}
    \partial\, \mathrm{STr}\, M^{-1} \bar\partial M =
    \bar\partial\, \mathrm{STr}\, M \partial M^{-1}
\end{equation}
is parity-invariant, a parity transformation reverses the orientation of the integration manifold $\Sigma$, thereby reversing the sign of the integral.

To appreciate the consequences of adding such a perturbation, we look at its expression in terms of the variables $\theta, \varphi$ of the effective action (\ref{eq:GFF}) for the $r=1$ theory:
\begin{equation}
    S_{\rm pert}^+ = \frac{\delta_+}{\mathrm{i}\pi} \int_\Sigma \partial\, \bar\partial\, (\varphi - \mathrm{i}\theta) = \frac{\delta_+}{2\pi} \int_\Sigma d^2z\, (- \Delta\varphi + \mathrm{i} \Delta\theta) ,
\end{equation}
where $\Delta = \partial_x^2 + \partial_y^2$ is the Laplacian of $\Sigma$. Now according to the equations of motion for the fields $\varphi, \theta$ one has $\Delta \varphi = 0$ and $\Delta \theta = 0$. However, the naive equation $\Delta \varphi = 0$ neglects the possible presence of vortex singularities $\varphi \sim \pm \ln | z - z_p |$. In fact, with the normalization chosen in (\ref{eq:S-rel}), $S_{\rm pert}^+$ counts the (algebraic) number of vortices. Thus, depending on the sign of the coupling $\delta_+$, the perturbation favors vortices of one charge and disfavors those of the opposite charge. In view of the parity-breaking effect of $\delta_+$, we propose the identification
\begin{equation}
    \delta_+ \sim \sigma_{xy} - \sigma_{xy}^\ast \,.
\end{equation}

To elucidate the practical consequences of the perturbation (\ref{eq:S-rel}), we need to renormalize it, taking into account any non-perturbative effects due to the vortex singularities in $M$. That is a task for the future. Whatever the outcome, it is clear by power counting that the perturbation (\ref{eq:S-rel}) is marginal, thereby supporting our suggestion that $1/\nu = 0$.

\section{Mean conductance}\label{sect:av-G}

A physical quantity readily accessible through experiments and numerical simulations is the electrical conductance. In numerical work (see e.g.\ \cite{KOK}), the standard system geometry to study is that of a cylinder of circumference $W$ and length $L$, with absorbing boundary conditions imposed at the two boundary circles. To obtain the (dissipative) conductance, one takes the trace of the square of an end-to-end Green's function or, assuming the Landauer-B\"uttiker formula, of the transmission matrix. In the presence of disorder, the conductance is a random variable with a distribution that widens with increasing $L$.

If we were able to perform an accurate field-theory calculation of the moments of the conductance on and off criticality, we would have a direct handle on our field-theory parameters and, in particular, we could read off the renormalization-group beta functions. Alas, that goal is too ambitious for now. What we can handle at present is the mean conductance, $G^\ast$, for a cylinder geometry in the scaling limit $L \to \infty$ (holding $W/L$ fixed), at the critical point of the IQH transition. In fact, we shall derive the formula
\begin{equation}\label{eq:G-sum}
    G^\ast = \frac{1}{\sqrt{\pi \tau}} \sum_{l \in \mathbb{Z} + 1/2} \mathrm{e}^{- l^2 \tau} = \frac{1}{\tau} \sum_{m \in \mathbb{Z}} (-1)^m \mathrm{e}^{- \pi^2 m^2 / \tau} \,, \quad \tau \equiv \frac{2\pi L}{n W} \,,
\end{equation}
valid (in the scaling limit of the critical system with ideal contacts) for any aspect ratio $L/W$. Note that the two expressions for the fixed-point conductance $G^\ast$ imply one another by Poisson summation. By specializing to the Ohmic regime $\tau \ll 1$ and matching to Ohm's law $G^\ast = \sigma_{xx}^\ast W / L$, one infers the value
\begin{equation}
    \sigma_{xx}^\ast = \frac{n}{2 \pi} = \frac{2}{\pi} \approx 0.6367 \quad (n = 4)
\end{equation}
for the fixed-point conductivity. We also note that the corrections (to Ohmic behavior) for the square geometry,
\begin{equation}
    G_{L=W}^\ast = \frac{2}{\pi} (1 - 2\, \mathrm{e}^{-2\pi} + \ldots) ,
\end{equation}
are too small to account for the lower value ($G_{L=W}^\ast = 0.57 \pm 0.02$) found in Ref.\ \cite{KOK}.

To start the derivation of Eq.\ (\ref{eq:G-sum}), we take the cylinder coordinates to be $x \in [0,L]$ and $y \in \mathbb{R} / (W \mathbb{Z})$. Thus the boundary circles are at $x = 0$ and $x = L$. Now according to Kubo linear-response theory, a conductance is a current-current correlation function. If $j_\parallel$ denotes the longitudinal (or $x$-) component of the current density operator (of a microscopic model in second quantization), our conductance $G$ is a ground-state expectation value
\begin{equation}\label{eq:Kubo}
    G = \left\langle \oint_0^W \!\!\!\! dy \, j_\parallel(x,y) \cdot \oint_0^W \!\!\!\! dy^\prime \, j_\parallel (x^\prime, y^\prime) \right\rangle .
\end{equation}
The two integrals are over any two cycles of constant longitudinal position. (By current conservation, the expression for $G$ is independent of the cycle positions $x$, $x^\prime$ as long as $0 < x < x^\prime < L$. The cycles may also be deformed within their homology classes.) Assuming a free-fermion ground state to arrive a reduced expression (a.k.a.\ the Kubo-Greenwood formula) in first quantization, one replaces the expectation $\langle j_\parallel (x,y) j_\parallel (x^\prime,y^\prime) \rangle$ by a product of retarded and advanced Green's functions,
\begin{equation*}
    \mathrm{G}^{\rm ret}(\mathbf{r},\mathbf{r}^\prime)\, \mathrm{G}^{\rm adv}(\mathbf{r}^\prime ,\mathbf{r}) , \quad \mathbf{r} = (x,y), \quad \mathbf{r}^\prime = (x^\prime,y^\prime) .
\end{equation*}
which are evaluated at the Fermi energy and decorated with derivatives ($\nabla_\parallel$ acting on one and $-\nabla_\parallel$ acting on the other factor; cf.\ (\ref{eq:KGref})) to form the longitudinal current.

What is our field-theory expression for the Kubo conductance (\ref{eq:Kubo}) averaged over the disorder? The answer (still using the symbol $G$, but now for the average conductance) is
\begin{equation}\label{eq:G-FT}
    G = \frac{1}{2} \left\langle \oint_0^W \!\!\!\! dy \, (\mathcal{J}_\parallel)_{\;1}^0 (x,y) \cdot \oint_0^W \!\!\!\! dy^\prime \, (\mathcal{J}_\parallel)_{\;0}^1 (x^\prime, y^\prime) \right\rangle ,
\end{equation}
where $\langle \ldots \rangle$ means the expectation value w.r.t.\ the functional integral with action (\ref{eq:def-actn}). We will now discuss the field-theory current $\mathcal{J}$ and overall factor $1/2$ in sequence.

While a first-principles derivation of the current $\mathcal{J}$ looks like a daunting task, we are assisted by the circumstance that $\mathcal{J}$ is essentially determined by symmetry. In fact, our current $\mathcal{J}$ corresponds to the conserved current of the underlying stationary disordered electron state at criticality, and as such it must be obtained as the linear response of the action functional (\ref{eq:def-actn}) w.r.t.\ axial transformations $M \mapsto \mathrm{e}^{X_L} M \mathrm{e}^{-X_R}$ with $-X_L = X_R \equiv X :$
\begin{equation}\label{eq:axial-J}
      S[\mathrm{e}^{-X} M \mathrm{e}^{-X}] = S[M] - \mathrm{i} \int \mathrm{STr} \, \mathcal{J} \wedge dX + \mathcal{O}(X^2) ,
\end{equation}
which are symmetries in the case of constant $X$. Such transformations leave the functional integration measure unchanged for any space-dependent $X$. Therefore, one has the equation of motion
\begin{equation}\label{eq:conserve}
    d\, \mathcal{J} = 0
\end{equation}
(as a Ward identity, valid under the functional integral sign and away from any operator insertions). From the definition (\ref{eq:axial-J}) one obtains the conserved current $\mathcal{J}$ as
\begin{equation}\label{eq:51}
    \mathcal{J} = \frac{J_L + J_R}{2\pi} \,,
\end{equation}
where $J_L$, $J_R$ are given by the expressions (\ref{eq:JL}, \ref{eq:JR}) for the CFT (\ref{eq:def-actn}). What appears in Eq.\ (\ref{eq:G-FT}) are the off-diagonal (even-odd and odd-even) components of $\mathcal{J}$; upon specializing to $r = 1$ we can label ``even'' and ``odd'' simply by $0$ and $1$.

Let us now explain the origin of the factor of 1/2 in Eq.\ (\ref{eq:G-FT}). (Without that factor, we would arrive at Ohm's law with an incorrect conductivity $\sigma_{xx}^\ast = n / \pi$.) The current $\mathcal{J}$ arises by bosonization of expressions that are quadratic in the bosonic and fermionic basic fields ($b_{\rm ret} \equiv b_+$, $f_{\rm adv} \equiv f_-$, etc.) of the SUSY formalism:
\begin{eqnarray}
    &&2\pi \mathcal{J}_{\ 1}^0 = (J_L + J_R)_{\; 1}^0 : \quad B_- F_+ - C_- G_+ = \mathrm{e}^{ \mathrm{i}\vartheta_1} b_+ f_- + \mathrm{e}^{- \mathrm{i}\vartheta_0} \bar{b}_- \bar{f}_+ \,, \cr &&2\pi \mathcal{J}_{\ 0}^1 = (J_L + J_R)_{\; 0}^1: \quad F_- B_+ - G_- C_+ = \mathrm{e}^{ -\mathrm{i}\vartheta_1} \bar{f}_- \bar{b}_+ - \mathrm{e}^{\mathrm{i}\vartheta_0} f_+ b_- \,.
\end{eqnarray}
The definitions of the linearly transformed fields $B_\pm$, $C_\pm$, $F_\pm$, $G_\pm$ are not reproduced here (for those, and for the {\it raison d'etre} of the phase factors $\mathrm{e}^{\mathrm{i} \vartheta_0}$ and $\mathrm{e}^{\mathrm{i} \vartheta_1}$, please be referred to our earlier paper \cite{CFT-IQHT}) since all we need for present purposes is that the product of Green's functions $\mathrm{G}^{\rm ret} \mathrm{G}^{\rm adv}$ appears \emph{twice} in the current-current correlator $\langle \mathcal{J}_{\ 1}^0 \mathcal{J}_{\ 0}^1 \rangle$, once from $\langle b_+ \bar{b}_+ \rangle \langle f_- \bar{f}_-\rangle$ and again from $-  \langle \bar{f}_+ f_+ \rangle \langle \bar{b}_- b_- \rangle$. This duplicity is brought about by rank reduction, $\widehat{\mathfrak{gl}}(2r|2r) \to \widehat{\mathfrak{gl}}(r|r)$, and the factor of $1/2$ in Eq.\ (\ref{eq:G-FT}) compensates for it.

\subsubsection{Perturbative calculation}\label{sect:G-pert}

Working with the $r=1$ theory, we now calculate the conductance $G^\ast$ in leading order of a loop (or $1/n$) expansion, at the RG-fixed point (\ref{eq:def-actn}) and for the cylinder geometry with absorbing boundary conditions at $x=0$ and $x=L$. First of all, we fix some constant isomorphism $I : \; \mathbb{C}_L^{1|1} \to \mathbb{C}_R^{1|1}$ as a reference map. Then, parameterizing the field as $M = \mathrm{e}^{X_L /2} I\, \mathrm{e}^{-X_R/2}$ with $X = X_L = - X_R\,$, we have the tree-level action
\begin{equation}
    S_{\rm tree} = \frac{1}{2} \int d^2z \; \mathrm{STr}\,X \left(\frac{-n\Delta}{4\pi} \right) X ,
\end{equation}
with $d^2z = |dx \wedge dy|$ and the Euclidean Laplace operator $\Delta = \partial_x^2 + \partial_y^2\,$. The quadratic contribution from the $\gamma$-deformation term has been omitted from $S_{\rm tree}\,$, as it does not affect in leading order the $\mathcal{J}_{\ 1}^0 \mathcal{J}_{\ 0}^1$ correlator of interest. {}From Eq.\ (\ref{eq:51}) the expression for the $01$-component of the total current through the cycle $x = \mathrm{const}$ is
\begin{equation}
    \oint dy\, (\mathcal{J}_\parallel )_{\; 1}^0 (x,y) = \frac{\mathrm{i} n}{2\pi} \oint dy\, \frac{\partial}{\partial x} X_{\; 1}^0 (x,y) + \mathcal{O}(X^2) .
\end{equation}
By calculating a Gaussian integral, we obtain the tree-level conductance as
\begin{equation}\label{eq:Gtree}
    G_{\rm tree} = \frac{1}{2} \, \left( \frac{n}{2\pi} \right)^2 \oint dy \oint dy^\prime \, \frac{- \partial^2}{\partial x\, \partial x^\prime} \left( \frac{- n \Delta}{4\pi} \right)^{-1}(x,y;x^\prime,y^\prime) .
\end{equation}
The inverse of the Laplacian (for $0 \leq x < x^\prime \leq L$) has the expression
\begin{equation}\label{eq:ffGFn}
    (-\Delta)^{-1}(x,y;x^\prime,y^\prime) = \sum_{k \in 2\pi\mathbb{Z}/W} \mathrm{e}^{\mathrm{i} k (y-y^\prime)} \; \frac{\sinh(kx)\, \sinh(k(L-x^\prime))}{k W \sinh(k L)} \,,
\end{equation}
where the summand for $k = 0$ is to be understood by the l'Hospital rule. To check that this is the correct free-field Green's function to use, notice that it vanishes at the boundaries ($x = 0$ and $x^\prime = L$) and behaves near the diagonal as $\Delta^{-1}(\mathbf{r},\mathbf{r}^\prime) = (2\pi)^{-1} \ln |\mathbf{r} - \mathbf{r}^\prime|$.

Only the $k = 0$ term in the sum of Eq.\ (\ref{eq:ffGFn}) survives the integration over transverse cycles in Eq.\ (\ref{eq:Gtree}). Thus we obtain the simple result
\begin{equation}\label{eq:Gtree2}
    G_{\rm tree} = \frac{1}{2} \left( \frac{nW}{2\pi} \right)^2 \frac{- \partial^2}{\partial x \,\partial {x^\prime}} \frac{x(L-x^\prime)}{LW} \, \frac{4\pi}{n} = \frac{n}{2\pi}\, \frac{W}{L} \,,
\end{equation}
which has the form of Ohm's law with fixed-point conductivity $\sigma_{xx}^\ast = n / 2\pi$. Away from the RG-fixed point, there are certainly corrections to the tree-level result (\ref{eq:Gtree2}). Nonetheless, right at the fixed point the result (\ref{eq:Gtree2}) holds exactly, by the following line of reasoning.

\subsubsection{Exact calculation}\label{sect:G-exact}

For the calculation of the first moment of the conductance distribution (as opposed to higher moments or cumulants thereof) it suffices to consider $r = 1$, as before. Taking advantage of the technical simplicity afforded by small matrix dimension, we parameterize the $2 \times 2$ supermatrix field $M$ as
\begin{equation}\label{eq:gauss}
    M = \begin{pmatrix} M^{0}_{\;\; 0} &M^{0}_{\;\; 1} \cr M^{1}_{\;\; 0} &M^{1}_{\;\; 1} \end{pmatrix} = \begin{pmatrix} 1 &0 \cr \eta &1 \end{pmatrix} \begin{pmatrix} \mathrm{e}^\varphi &0 \cr 0 &\mathrm{e}^{\mathrm{i}\theta} \end{pmatrix}
    \begin{pmatrix} 1 &\xi \cr 0 &1 \end{pmatrix} .
\end{equation}
What makes this parametrization especially useful is that the fixed-point action functional (\ref{eq:def-actn}) has a quadratic dependence on the fermionic fields $\xi$ and $\eta$. Yet, there is still the complication that $\xi$ and $\eta$ appear in the current operators $\mathcal{J}_\parallel$ of the correlation function (\ref{eq:G-FT}). Hence, to do the Gaussian integral over these fields, it is convenient to re-express the product of current operators as
\begin{equation}\label{eq:replace}
    (\mathcal{J}_\parallel)_{\ 1}^0 (\mathcal{J}_\parallel)_{\ 0}^1 \longrightarrow - {\textstyle{\frac{1}{2}}} (\mathcal{J}_\parallel)_{\ 0}^0 (\mathcal{J}_\parallel)_{\ 0}^0 + {\textstyle{\frac{1}{2}}} (\mathcal{J}_\parallel)_{\ 1}^1 (\mathcal{J}_\parallel)_{\ 1}^1 \,.
\end{equation}
(The re-expressed correlation function is identical to the previous one by the target-space supersymmetry of the formalism). Another useful trick is to push the cycles $x = \mathrm{const}$ and $x^\prime = \mathrm{const}$ all the way to the boundaries of system ($x \to 0$ and $x^\prime \to L$) by invoking current conservation. The current-current correlation function then becomes (a second derivative of) the partition-function integral with twisted boundary conditions. By the replacement (\ref{eq:replace}) the twisting is
\begin{equation}\label{eq:bc}
    \varphi(x=0,y) = 0 \to \varphi(0,y) = 2 \varepsilon, \quad
    \varphi(x^\prime=L,y^\prime) = 0 \to \varphi(L,y^\prime) = 2 \varepsilon^\prime
\end{equation}
for the first term on the r.h.s.\ of (\ref{eq:replace}), and similar for the second term (with $\varphi$ replaced by $\theta$). In the next paragraph, we describe that trick in more detail.

Focusing on the first term on the right-hand side of (\ref{eq:replace}), consider the discontinuous $00$-generator field $X_{\varepsilon,\varepsilon^\prime}$ defined by
\begin{equation}
    X_{\varepsilon, \varepsilon^\prime} = (\varepsilon\, \chi_{<x} + \varepsilon^\prime \chi_{>x^\prime}) \begin{pmatrix} 1 &0 \cr 0 &0 \end{pmatrix} ,
\end{equation}
where $\chi_{<x}$ and $\chi_{>x^\prime}$ are the characteristic functions for the two collar regions delineated on the cylinder by the two cycles of constant $x$ and $x^\prime$ in Eq.\ (\ref{eq:G-FT}). Then the functional integral $\int \mathrm{e}^{-S}$ with
\begin{equation}\label{eq:trick}
    S = S_{n,\gamma}^{\rm WZW}[\mathrm{e}^{X_{\varepsilon,\varepsilon^\prime}} M\,
    \mathrm{e}^{X_{\varepsilon,\varepsilon^\prime}} ]
\end{equation}
generates our current-current correlator by Taylor expansion as the coefficient of the product $\varepsilon\, \varepsilon^\prime$, by virtue of (\ref{eq:axial-J}) and the jump of the characteristic functions. Next, we do the step of pushing the current integration cycles to the boundaries $x = 0$ and $x^\prime = L$. In the final step, we absorb the left and right factors $\mathrm{e}^{X_{\varepsilon, \varepsilon^\prime}}$ in Eq.\ (\ref{eq:trick}) into the integration variable $M$, thereby twisting the boundary conditions as in (\ref{eq:bc}).

By all these manipulations, the integral over the fermionic fields $\xi$ and $\eta$ becomes a straight Gaussian integral, giving a log-determinant correction \cite{CFT-IQHT}
\begin{equation}\label{eq:log-det}
    - \ln \mathrm{Det} \big( - \mathrm{e}^{-\lambda /2} \, \partial_z \, \mathrm{e}^{\lambda} \, \partial_{\bar z} \, \mathrm{e}^{ -\lambda /2} \big) , \quad \lambda \equiv \varphi - \mathrm{i} \theta \,,
\end{equation}
to the effective action functional. Luckily, that log-determinant can be computed exactly. Its effect is to cancel the $\gamma$-deformation term and add instead a term $(\varphi - \mathrm{i} \theta) \mathcal{R} Q$ with ``background charge'' $Q = - 1$ to the Lagrangian:
\begin{equation}\label{eq:GFF}
    S_{n,\gamma=1}^{\rm eff} = \frac{\mathrm{i} n}{4\pi} \int_\Sigma \big( \partial\varphi \wedge \bar\partial \varphi + \partial\theta \wedge \bar\partial\theta \big) - \frac{Q}{8\pi} \int_\Sigma (\varphi - \mathrm{i}\,\theta) \mathcal{R} .
\end{equation}
The 2-form $\mathcal{R}$ is the curvature of the Riemann surface $\Sigma$. In the case at hand (of a cylinder $\Sigma$ with flat-space geometry), $\mathcal{R}$ is localized at the two system boundaries: one may imagine that the boundary conditions (\ref{eq:bc}) compactify the cylinder to a two-sphere with singular curvature at the two points ($x=0$ and $x=L$) of compactification.

It remains to work out the partition functions for the Gaussian free field $\varphi$ and the compact boson $\theta$, for our cylinder with twisted boundary conditions (\ref{eq:bc}). This computation can be done by the transfer matrix technique (or radial quantization). The result is in (\ref{eq:G-sum}), where we note that the sum over integers $l \in \mathbb{Z} + 1/2$ accounts for the periodicity $\theta \sim \theta + 2\pi$, and the shift by $1/2$ is caused by the background charges at $x = 0, L$.

\section{Outlook}

Past numerical work on the integer quantum Hall transition failed to reach conclusive agreement on the numerical values of the critical exponent $\nu$ for the localization length and the leading irrelevant exponent $y$. Our prediction \cite{CFT-IQHT} from the conformal field theory (CFT) describing the scaling limit of the transition was that, strictly speaking, $\nu^{-1} = 0 = y$, as the CFT admits nothing but current-current perturbations that are marginal. We proposed that the numerics sees transients due to terms in the renormalization-group (RG) beta functions which are nonlinear in the couplings of these marginal perturbations. In the present article, we investigated that proposal in detail. We found that one natural candidate for the marginally irrelevant perturbation can be ruled out on phenomenological grounds. The remaining candidate has the unorthodox feature that its coupling must be imaginary in order for the RG flow to be attracted to the CFT. In view of this inconclusive situation, there is now a strong need for a complete derivation of the CFT starting from a disordered electron model such as the network model of Chalker \& Coddington \cite{Chalker}.

Turning to the bigger picture, a major challenge to our theoretical scenario is posed by a variant of the Chalker-Coddington network model, namely the model with two channels per link and $\mathrm{SU}(2)$ disorder (a.k.a.\ class $C$, or ``spin'' quantum Hall effect). We believe that the RG-fixed point field theory of that model, some properties of which are known exactly, is a CFT very similar to the one reviewed here. The same phenomena of spontaneous symmetry breaking and rank reduction are expected to take place there.

\bigskip\noindent{\bf Acknowledgments.} The author benefitted from discussions with M.S.\ Foster and I.A.\ Gruzberg. He thanks A.A.\ Saberi and H.\ Dashti-Naserabadi for collaboration related to the present work.


\begin{thebibliography}{99}
\bibitem{footnote-A} For the case of antiferromagnetic spin chains, such an argument was originally given by Affleck \cite{Affleck}; it was adapted to the present situation in \cite{CFT-IQHT}.
\end{thebibliography}
\end{document}